\documentclass[a4paper,11pt]{article}
\pdfoutput=1 
\usepackage{jinstpub} 
\usepackage{siunitx}
\usepackage{gensymb}

\title{\boldmath Simulation Study of the MPD Performance for Dielectron Measurements at NICA}

\author[a, 1]{Sudhir Pandurang Rode\note{Corresponding author.}}

\author[b]{Itzhak Tserruya}

\author[a,d]{Victor Riabov}

\author[c]{Yonghong Wang}
\author[c]{Chi Yang}

\affiliation[a]{Veksler and Baldin Laboratory of High Energy Physics, Joint Institute for Nuclear Research, Dubna, 141980, Moscow region, Russian Federation}
\affiliation[b]{Weizmann Institute of Science, Rehovot, 76100, Israel}
\affiliation[c]{Institute of Frontier and Interdisciplinary Science, Shandong University, Qingdao, 266237, China}
\affiliation[d]{National Research Nuclear University MEPhI (Moscow Engineering Physics Institute), Moscow, 115409, Russian Federation}

\emailAdd{sudhir@jinr.ru}

\abstract{The Multi-Purpose Detector (MPD) experiment at the Nuclotron-based Ion Collider fAcility (NICA) is designed to investigate strongly interacting matter at high net-baryon density through a broad program of heavy-ion measurements. Dielectrons constitute one of the key probes in this program, with contributions emitted throughout the entire space-time evolution of the collision. Their experimental measurement is, however, challenged by a large combinatorial background arising primarily from incompletely reconstructed photon conversions and Dalitz decays of light neutral mesons. In this work, we study the expected performance and capabilities of the MPD experiment for dielectron measurements using simulated minimum-bias $^{209}$Bi+$^{209}$Bi collisions at $\sqrt{s_{NN}}$ = 9.2 GeV. Electron identification is performed using the combined information from the time projection chamber (TPC), the time-of-flight (TOF) detector and the electromagnetic calorimeter (ECal). A previously developed multilayer-perceptron classifier is used for electron identification, providing a substantial increase in electron detection efficiency relative to sequential one-dimensional selections while preserving an electron-sample purity close to unity. To address the dominant sources of combinatorial background, a pair-analysis strategy is developed that exploits partially reconstructed electron tracks. The method combines the pair opening angle, the TPC dE/dx signal, and an approximate reconstruction of the pair invariant mass to tag tracks from photon conversions and $\pi^0$ Dalitz decays at the pair level. In the mass interval 0.2 $< m_{ee}<0.7 $ GeV/$c^2$, the method improves the signal-to-background ratio by a factor of about 3.5 with the current track reconstruction algorithm. The study demonstrates the strong potential of the MPD experiment for dielectron measurements at NICA energies and highlights the importance of further improvements in low-$p_{\rm T}$ track reconstruction.}


\begin{document}
\maketitle
\flushbottom

\section{Introduction}
Understanding the properties of strongly interacting matter at high net-baryon density remains one of the central goals of relativistic heavy-ion physics. The Nuclotron-based Ion Collider fAcility (NICA) has been designed to provide high-luminosity collisions of heavy ions in the nucleon-nucleon center-of-mass energy range of $\sqrt{s_{NN}} = 4$--$11$ GeV~\cite{Sissakian:2009zza}. 
In this energy range, heavy-ion collisions are expected to produce net-baryon-dominated matter, as opposed to the meson-dominated matter produced at the higher energies of the SPS, RHIC, and LHC. Model calculations indicate that the net-baryon density at freeze-out reaches a maximum at a collision energy of $\sqrt{s_{NN}} \approx 7.5$ GeV~\cite{Randrup:2006nr}.
NICA will enable systematic studies of nucleus–nucleus interactions over a broad range of colliding species and energies, offering a complementary approach to past and ongoing programs aimed at exploring dense QCD matter at higher collision energies. 
At the moment of writing, the NICA collider is in the commissioning phase.  

The MPD (Multi-Purpose Detector) is the flagship heavy-ion experiment of NICA. MPD is a large-acceptance apparatus optimized for measurements of hadrons and electromagnetic probes in heavy-ion collisions. 
The MPD physics program is broad, targeting fundamental questions such
as the search for a possible first-order phase transition from hadronic to partonic matter and a critical point in the QCD phase diagram, as suggested by several theoretical approaches~\cite{Bzdak2020}.
Furthermore, MPD will search for the onset of the deconfinement and chiral symmetry restoration phase transitions~\cite{MPD:2022qhn}.

Dielectron measurements are sensitive to these questions, making them particularly powerful probes. In particular, they provide direct access to the thermal radiation emitted
throughout the entire evolution of the collision, including the radiation emitted in the early stage of the quark-gluon plasma phase (if it is formed at the NICA energies) as well as the radiation from the subsequent hadron-gas phase~\cite{Tserruya:2009zt}. 
Although dileptons have been measured and studied extensively in relativistic heavy-ion collisions at the high energies available at the SPS and RHIC~\cite{CERES:1995, CERESNA45:1997tgc, CERESNA45:2002gnc, CERES:2005uih, CERES:2006wcq,
NA60:2006ymb, NA60:2008,  NA60:2009, NA60:2009-2, STAR2014, STAR2023, STAR:2024bpc, PHENIX:2016}, 
measurements at lower energies remain limited.  Results are available from DLS and HADES near $\sqrt{s_{NN}} = 2.4$ GeV 
\cite{DLS:1997kbk, HADES2019} and from the STAR beam energy scan \cite{STAR_HP24}. The MPD will substantially extend the dielectron measurements in the low-energy region. It will
allow systematic coverage with a single detector and a common analysis in the high-net-baryon-density region of the QCD phase diagram. 

The measurement of dielectrons in heavy-ion collisions is notoriously difficult. 
The dielectron signal is embedded in a large combinatorial background arising from incompletely reconstructed photon conversions and light-meson Dalitz decays, together with residual misidentified hadrons. 
Achieving meaningful physics results requires excellent charged-particle tracking in a high-multiplicity environment, robust primary and secondary vertex reconstruction, efficient electron identification and strong hadron rejection over a broad kinematic range. 

This work presents a dedicated simulation study of the MPD performance for dielectron measurements at NICA energies, building on the preliminary feasibility study reported as part of the broader MPD physics-performance survey in ref.~\cite{MPD:2025jzd}. That study used sequential one-dimensional electron-identification selections and introduced a preliminary low-mass pair-rejection procedure involving fully and partially reconstructed electron candidates. In the present analysis, the MLP-based electron-identification method developed in ref.~\cite{Rode:2026mmj} is employed as a more efficient analysis ingredient. In addition, the pair-rejection procedure is developed, introducing separate selections according to the detector information available for the partially reconstructed partner and validating these selections using generator-level information. The resulting performance is quantified in terms of electron efficiency and purity, background rejection, signal retention, signal-to-background ratio, and statistical significance.

The paper is organized as follows: In section~\ref{sec2}, we briefly introduce the MPD apparatus and the detector subsystems relevant to electron measurements. Section~\ref{sec_simul} describes the simulated event samples, detector response, and reconstruction.
Section~\ref{sec_eid-perf} presents the MPD electron-identification performance showcasing the benefit of using machine learning. 
Section~\ref{sec_challenges} describes the principal challenges in $e^{+}e^{-}$ pair reconstruction. 
Section~\ref{pair_analysis} introduces the tracklet-assisted strategy for suppressing the combinatorial background. 
Section~\ref{results} presents the results, and a summary of the study is given in section~\ref{summ}.
 
\section{MPD apparatus and particle identification}
\label{sec2}
\subsection{Detector layout}

\begin{figure*}
\centering
\includegraphics[scale=0.3]{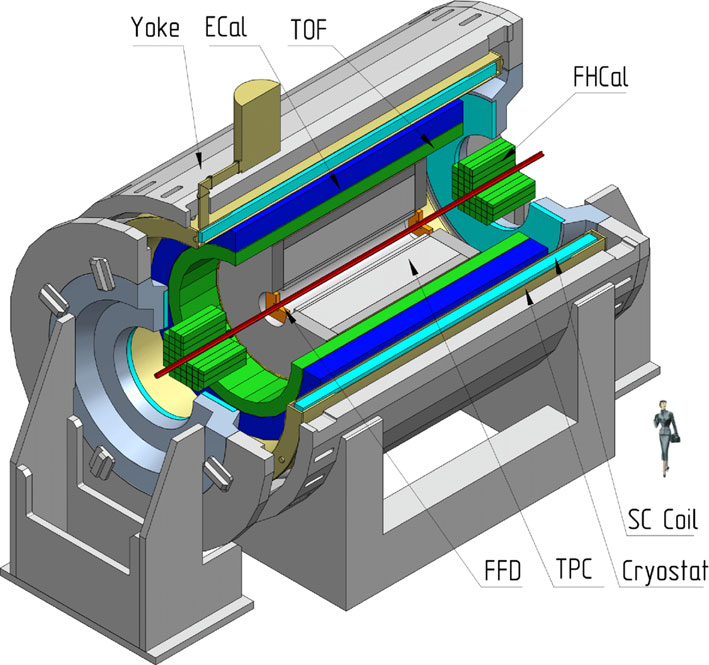}
\caption{Schematic view of the MPD apparatus. The central barrel subsystems ordered from inside to outside are TPC, TOF and ECal and the forward subsystems are FFD and FHCal.}
\label{fig_mpd}
\end{figure*}

The MPD is designed to measure and identify charged hadrons, electrons, and photons over a wide range of momentum and rapidity. A schematic view of the MPD apparatus as planned for the first stage of operation is shown in figure~\ref{fig_mpd}. A detailed description is provided in ref.~\cite{MPD:2022qhn}. Here, we briefly present only the subsystems relevant to the measurement of electrons. 

The MPD experiment is built inside
a superconducting solenoid providing a nominal magnetic field of  0.5~T.  
The detector has full azimuthal coverage and a pseudorapidity acceptance of approximately
$|\eta| < 1.5$ for charged particles. The present electron analysis is restricted to $|\eta| < 1.0$, where the relevant subsystem coverage and reconstruction performance are sufficiently uniform.

The central detector region hosts the main tracking and particle
identification subsystems: a time projection chamber (TPC),  a time of flight (TOF) detector and an electromagnetic calorimeter (ECal). The setup also includes two subsystems in the forward rapidity region, a fast forward detector (FFD) for vertex and start-time determination and a forward hadron calorimeter (FHCal), used for centrality and event-plane determination.  

\subsection{TPC tracking and $dE/dx$ measurement}
The TPC is the main tracking detector of the MPD experiment. It provides three-dimensional reconstruction of charged-particle trajectories and momenta~\cite{Vereschagin:2020rgn} with a relative momentum resolution of $\sim2\%$ at p$_T$ of 
1 GeV/c~\cite{MPD:2022qhn}.
The detector is cylindrical with a length of 3.4~m, and inner and outer radii of 0.34~m and 1.4~m, respectively.  The TPC sensitive volume is divided into two drift regions by a central high-voltage electrode. The drift gas is a mixture of 90\% argon and 10\% methane (P10), chosen for its good drift properties and stable detector operation.
The electric drift field is approximately 140~V/cm. 

Ionization electrons produced by charged particles drift toward the end-cap readout chambers where they are amplified and detected by multi-wire proportional chambers with cathode pad readout.
Each endcap contains 12 readout chambers (ROC), resulting in a total of 24 ROCs for the full detector. 
Each ROC has a trapezoidal pad plane segmented into approximately $4000$ readout pads arranged in 53 pad rows (perpendicular to the radial direction), with 
\begin{figure*}
\centering
\includegraphics[scale=0.23]{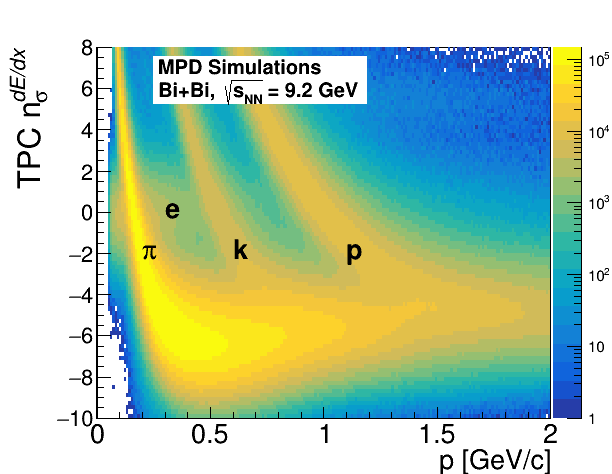}
\includegraphics[scale=0.23]{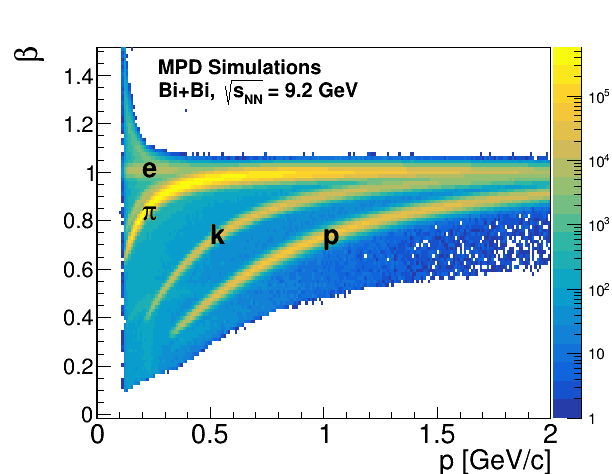}
\includegraphics[scale=0.23]{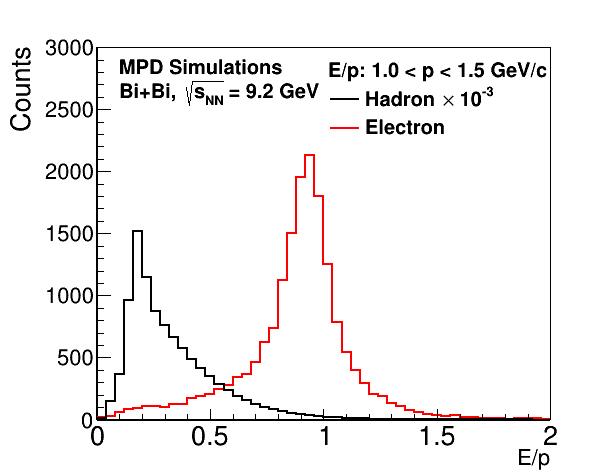}
\caption{Detector responses for reconstructed tracks in simulated minimum-bias Bi+Bi collisions at $\sqrt{s_{NN}}$=9.2 GeV. Left: TPC $dE/dx$ signal, expressed in terms of the number of sigmas of the deviation from the expected electron energy loss as a function of the momentum p. Middle: TOF velocity $\beta$ as a function of p. Right: $E/p$ ratio of electrons and hadrons in the ECal. The overwhelming yield of hadrons is scaled by a factor of $10^{-3}$ to make the electrons visible.}
\label{fig_pid}
\end{figure*}  
two pad sizes ($5\times12$~mm$^2$ in the inner region and $5\times18$~mm$^2$ in the outer region) in order to maintain uniform occupancy and position resolution. 

In addition to tracking, the TPC provides particle identification via measurements of the specific ionization energy loss $dE/dx$. 
The energy loss, $dE/dx$, is calculated as the truncated mean of the lowest 70\% of the cluster-charge measurements assigned to a track; the highest 30\% of the measurements are discarded to reduce the influence of the Landau tail~\cite{Gertsenberger:2016llj}. The TPC allows separation of charged pions from kaons up to momenta of $\sim$~0.7 GeV/c and kaons from protons up to $\sim$~1.1 GeV/c, both corresponding to a 2$\sigma$ level separation \cite{MPD:2022qhn}.

The left panel of figure~\ref{fig_pid} shows the $dE/dx$ signal expressed in terms of the number of sigmas, i.e. the deviation of the measured energy loss $(dE/dx)_{meas}$  from the expected electron energy loss $\mu_e$ in the TPC gas volume normalized to its resolution $\sigma_e$, i.e. 
\begin{equation}
    n_{\sigma}^{dE/dx} = ((dE/dx)_{meas} - \mu_e) / \sigma_e
\end{equation}
In this representation the electrons populate the region around $n_{\sigma}^{dE/dx}$=0.

\subsection{Time-of-Flight detector}

The TOF system provides particle identification by measuring the flight time of charged particles along the reconstructed track from the primary vertex to the matched TOF hit. 
The detector, based on the multigap resistive plate chamber (MRPC) technology, is arranged in a cylindrical barrel geometry. It is located around the TPC and consists of 14 sectors. Each sector has two modules and every module contains 10 MRPCs with 24 readout strips per MRPC. Signals are read from both ends of each strip. The complete system therefore comprises 28 modules, 280 MRPCs and 13440 readout channels. 
The mechanical support structure and the boundaries between adjacent modules introduce inactive regions  in the TOF acceptance that amount to $\sim$10\%.
The TOF system provides both time and coordinate measurements with an accuracy of $\sim$~80 ps and $\sim$~0.5 cm, respectively \cite{Babkin2018TOF}. The quoted 80 ps corresponds to the intrinsic resolution of the TOF detector. The collision start time $t_{0}$ is supplied by the FFD, which is designed to provide a start-time resolution better than 50 ps~\cite{MPD:2022qhn}. In the present simulation, the TOF hit time is smeared according to the intrinsic TOF resolution, while $t_{0}$ is additionally smeared according to the centrality-dependent FFD resolution. The middle panel of figure~\ref{fig_pid} displays the particle velocity $\beta$ measured by the TOF versus the momentum $p$ measured by the TPC.
Because electrons reach relativistic velocities already at very low momentum, they populate the region near $\beta \approx 1$, while heavier hadrons exhibit smaller velocities in the same momentum range, resulting in an efficient separation of electrons, pions, kaons and protons at low momentum.

\subsection{Electromagnetic calorimeter}

The ECal provides an independent measurement of the position and energy of electromagnetic showers induced by electrons and photons.

The ECal is a sampling calorimeter of the Shashlyk type, made of alternating layers of lead absorber and plastic scintillator plates~\cite{Semenov:2020glg}. 
Geometrically, it consists of 25 sectors or 50 half-sectors forming a 6 m long cylindrical shell located outside the TOF detector. In the transverse plane, each sector covers an azimuthal angular range of 14.4$^\circ$. 

A half-sector contains 48 calorimeter modules, arranged as eight modules of different types along the longitudinal direction and six along the azimuthal direction. The modules point toward the interaction point, forming a projective geometry. This design ensures that particles produced in the collision traverse approximately the same amount of material, independently of their angle, thereby improving the uniformity of the calorimeter response and reducing shower leakage.
Each module consists of 16 towers that are glued together. Each tower has a 40 × 40 mm$^2$ transverse cross-section and is composed of a lead-scintillator sandwich that contains 210 tiles of Pb (0.3 mm thick each) interleaved with 210 tiles of plastic scintillator (1.5 mm thick each), resulting in a total thickness of about 11 radiation lengths. 

Electron identification using the calorimeter is based on the ratio $E/p$ between the measured energy $E$ in the calorimeter and the momentum $p$ reconstructed in the TPC.

Electrons whose electromagnetic showers are largely contained in the calorimeter produce a characteristic peak around $E/p \approx 1$, whereas hadrons typically deposit only a fraction of their energy and populate the lower $E/p$ region. This is illustrated in the right panel of figure~\ref{fig_pid} that shows the ECal response to electrons and hadrons separately, in the momentum interval 1.0 $<$ p $<$ 1.5 GeV/c. The $E/p$ observable is particularly valuable for electron identification  at high momentum where the separation power of the $dE/dx$ and TOF measurements decreases.

The ECal also provides time-of-flight information. In the simulations reported here, a conservative time resolution of 500 ps is used, approximately twice the expected value. Although this resolution is inferior to that provided by the TOF detector,  the ECal timing information is still useful as a supplement in regions where the TOF information is unavailable.
  
\section {Simulations and reconstruction} 
\label{sec_simul}

The study presented here is carried out on a sample of 32.6 million minimum-bias $^{209}\rm Bi+^{209}\rm Bi$ collisions at a nucleon-nucleon center-of-mass energy of $\sqrt{s_{NN}} = 9.2$ GeV generated with the UrQMD version 3.4 event generator~\cite{Bass:1998ca,Bleicher:1999xi}. The impact parameter b was sampled uniformly in $b^2$ over the interval $b$ = 0-15 fm.  

\begin{figure*}
\begin{center}
\includegraphics[scale=0.3]
{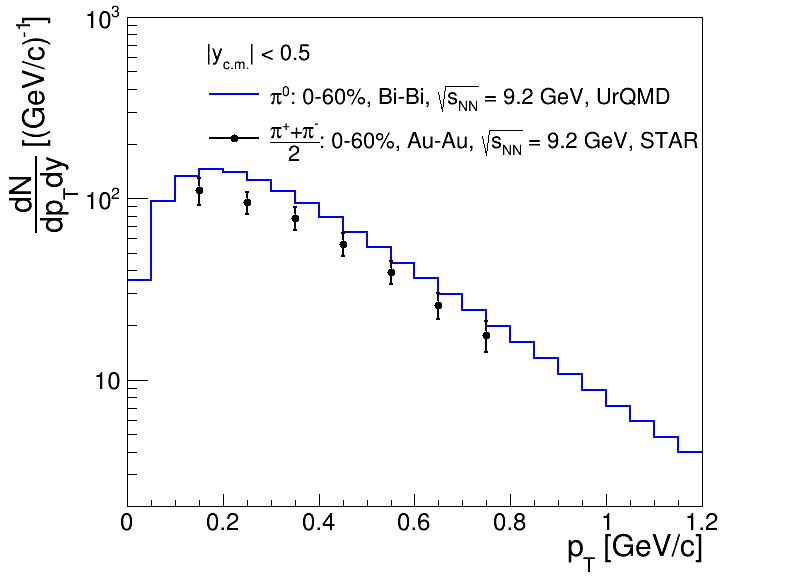}
\caption{Comparison of the $\pi^{0}$ transverse-momentum spectrum from UrQMD Bi+Bi collisions with the average charged-pion spectrum, 1/2($\pi^-+ \pi^+$), measured by STAR in Au+Au collisions at $\sqrt{s_{NN}}=$9.2 GeV ~\cite{STAR:2009sxc}. The same centrality and rapidity intervals are used.}
\label{fig3}
\end{center}
\end{figure*}

The hadronic final state produced by UrQMD provides the underlying event and the principal sources that contribute to the combinatorial dielectron background. Figure~\ref{fig3} compares the $\pi^0$ transverse-momentum distributions from UrQMD with the average charged-pion spectrum, 1/2($\pi^-+ \pi^+$), measured by STAR in Au+Au collisions at the same collision energy~\cite{STAR:2009sxc}. The average charged-pion spectrum is used as a proxy for the neutral-pion spectrum. The simulation and data are compared in the same centrality and rapidity intervals. UrQMD reproduces the magnitude and shape of pion production reasonably well, indicating that UrQMD provides a reasonable estimate of the pion-driven combinatorial background. 

A sample of 32.6 million collisions does not provide sufficient statistical precision for the rare dielectron decays of light vector mesons. To have a visible signal in this relatively small sample, artificially enhanced branching ratios are used for the following rare electron sources: the decays $\omega\rightarrow e^{+}e^{-}$, 
$\omega\rightarrow \pi^{0 }e^{+}e^{-}$, 
$\rho\rightarrow e^{+}e^{-}$, 
$\phi\rightarrow \eta e^{+}e^{-}$ and 
$\phi\rightarrow e^{+}e^{-}$, are enhanced by a factor of 20. Additionally, the $\eta$ Dalitz decay, $\eta\rightarrow e^{+}e^{-} \gamma $, is enhanced by a factor of 5.
These enhancements are introduced only to improve the statistical precision of the simulated samples. They have a negligible effect on the inclusive reconstructed electron multiplicity, which is dominated by photon conversions and $\pi^{0 }$ Dalitz decays, whose rates are not artificially enhanced.
Inverse enhancement factors are applied to restore the physical contributions of the oversampled sources when constructing  the final results of invariant mass spectra and integrated yields. Generator-level information from UrQMD is used to determine the downscaling factor for each pair according to the origins of its two tracks.

In addition, UrQMD does not reproduce well the yield, and in particular the spectral shape, of the vector meson decays into $e^+e^-$. Therefore, the UrQMD dielectron yields from these sources were weighted to the predictions of the PHSD model ~\cite{PHSD,PHSD2} which provides a reasonable description of the dielectron spectra measured by STAR over the entire RHIC energy range ~\cite{Jorge:2025wwp}. For a source $k$, the weight $w$ is only a function of the dielectron mass $m$, i.e. $w_k(m) = N_k^{PHSD}(m) / N_k^{UrQMD}(m) $. This procedure corrects the shape and yield of the dielectron mass spectrum and does not alter the generated pair kinematics. We have compared the transverse-momentum distributions of the corresponding sources in UrQMD and PHSD and find them to be similar, indicating that the mass-only reweighting does not introduce a significant mismatch in the acceptance-sensitive p$_T$ distributions. Residual differences are not corrected by this procedure and constitute a model limitation of the present performance study.
  
The generated particles are transported and tracked through the detector material using GEANT4  \cite{GEANT} and the geometrical configuration corresponding to the first-stage of the MPD setup described in section~\ref{sec2}. The simulation uses the MPD nominal magnetic field of  $B$ = 0.5 T implemented using a uniform-field approximation.

The response of the detector subsystems and the global tracking are performed using the MpdRoot code, which is the software framework of the MPD collaboration~\cite{mpdroot}. Charged particle tracks in the TPC and the primary vertex are reconstructed using the Kalman filtering technique~\cite{Fruhwirth:1987fm,Luchsinger:1992np}. The TPC pattern recognition constructs track candidates from the inner to the outer pad rows, associates clusters with the candidates, and performs a helical track fit to determine their momentum. Further details on MPD tracking are given in ref.~\cite{Gertsenberger:2016llj}. The primary-vertex resolution improves with the number of reconstructed tracks; the expected longitudinal vertex resolution is presented in figure 18 of ref.~\cite{MPD:2022qhn}. The minimum-bias event sample spans a broad multiplicity range, representative distributions of which are shown for reconstructed Bi+Bi events in figure 40 of ref.~\cite{MPD:2022qhn}.

 \begin{table*}[ht]\centering
\vglue4mm
\setlength{\tabcolsep}{10pt} 
\renewcommand{\arraystretch}{1.2} 
\begin{tabular}{cc} \hline\hline
Variable & Requirement \\
\hline\hline

$|\eta|$ & $<$ 1.0 \\
Number of TPC hits & $>$ 15 \\
DCA cut & $|n^{\rm DCA_{x,y,z}}_{\sigma}|$ $<$ 2.5 \\
TPC-TOF matching & $|n^{\rm d\phi}_{\sigma}|$ $<$ 3 and $|n^{\rm dz}_{\sigma}|$ $<$ 2 \\
TPC-ECal matching & $|n^{\rm d\phi}_{\sigma}|$ and $|n^{\rm dz}_{\sigma}|$ $<$ 3 \\
\hline\hline
\end{tabular}
\caption{Track quality and matching cuts used to characterize the fully reconstructed tracks in this analysis.}
\label{tabI}
\end{table*}

\begin{figure*}
\centering
\includegraphics[scale=0.23]{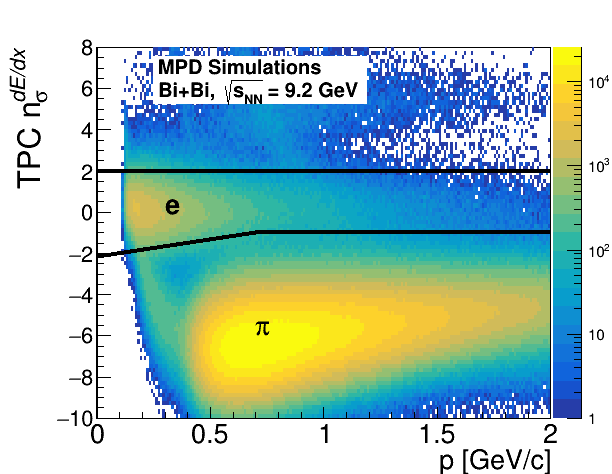}
\includegraphics[scale=0.23]{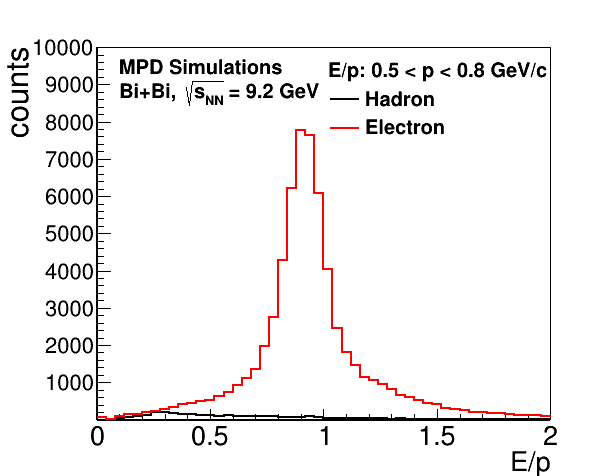}
\includegraphics[scale=0.23]{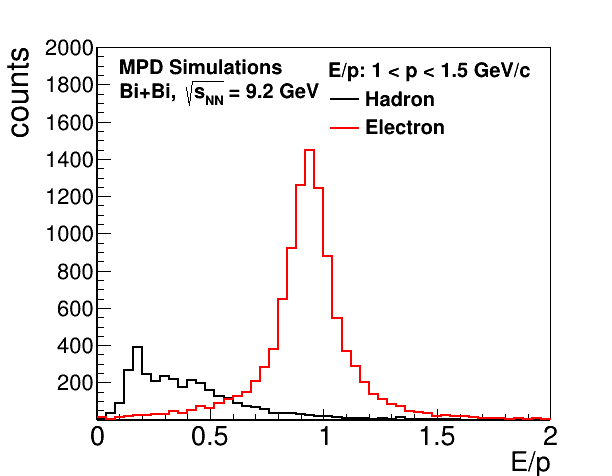}
\caption {Left: TPC $dE/dx$ signal, expressed in terms of the number of sigmas of the deviation from the expected electron energy loss after TOF eid selection as a function of momentum p. Middle and right panels: $E/p$ ratio of electrons and hadrons in the ECal in the momentum intervals 0.5 $<$ p $<$ 0.8 GeV/c and 1 $<$ p $<$ 1.5 GeV/c, after applying both the TPC $dE/dx$ selection, shown by the black lines in the left panel, and the TOF cut of $|n_{\sigma}^{\beta}| < 2$.}
\label{fig_pid2}
\end{figure*}

The reconstructed tracks are extrapolated back toward the beam axis, and a 2.5$\sigma$ cut is applied on the distance-of-closest approach (DCA) to the primary vertex. This cut removes a significant fraction of secondary tracks and, in particular, photon conversions as discussed below in section~\ref{sec_challenges}. The transverse and longitudinal DCA resolutions depend on the track momentum (see figure 18 of ref.~\cite{MPD:2022qhn}). The DCA cut in table~\ref{tabI} is therefore expressed in units of the corresponding track-by-track estimated resolution.

The TPC tracks are extrapolated to the outer detectors, TOF and ECal, and matched to the nearest hits applying the azimuthal and longitudinal matching requirements listed in table~\ref{tabI}. In the following, a fully reconstructed track refers to a TPC track matched to TOF and ECal and satisfying all quality and matching conditions listed in table~\ref{tabI}.
 
 \section{Electron identification performance}
 \label{sec_eid-perf}
The MPD electron identification (eid) performance is quantified in terms of the single-electron efficiency and the purity of the selected electron sample, in the pseudorapidity range 
$|\eta| < 1$. The efficiency is defined as the fraction of generated primary electrons that are reconstructed and pass the identification criteria. The purity is defined as the fraction of true electrons among the reconstructed tracks that satisfy the electron-identification criteria.

While the TPC, TOF and ECal, used separately,  exhibit limited electron identification, as shown in figure~\ref{fig_pid}, it is the combined effect of the three detectors that provides a powerful identification over the entire momentum range. This is illustrated in figure~\ref{fig_pid2}. 
The left panel shows the TPC $dE/dx$ signal after applying a TOF eid cut of $|n_{\sigma}^{\beta}| < 2$. Whereas the electron band is hardly visible in the stand-alone TPC $dE/dx$ signal (see left panel of figure~\ref{fig_pid}), it is  clearly seen and well separated after applying the TOF eid selection.
The middle and right panels show the $E/p$ ratio of electrons and hadrons in the ECal for two momentum intervals after adding the $dE/dx$ cut shown by the black lines in the left panel. The histograms reflect the electron and hadron abundances from the simulated UrQMD events. While the hadron contamination is negligible at low momentum, as illustrated in the middle panel, this is not so at higher momenta (right panel) demonstrating how the ECal is instrumental in obtaining a high-purity electron sample at high momentum. The electron $E/p$ peak in figure~\ref{fig_pid2} is slightly displaced from unity because the ECal response used in the present simulation has not been calibrated specifically to set the electron energy scale to $E/p$=1. The electron-identification selection is defined relative to the simulated electron response and is therefore not affected by this offset.

\subsection{Electron detection efficiency}
\begin{figure*}
\centering
\includegraphics[scale=0.3]{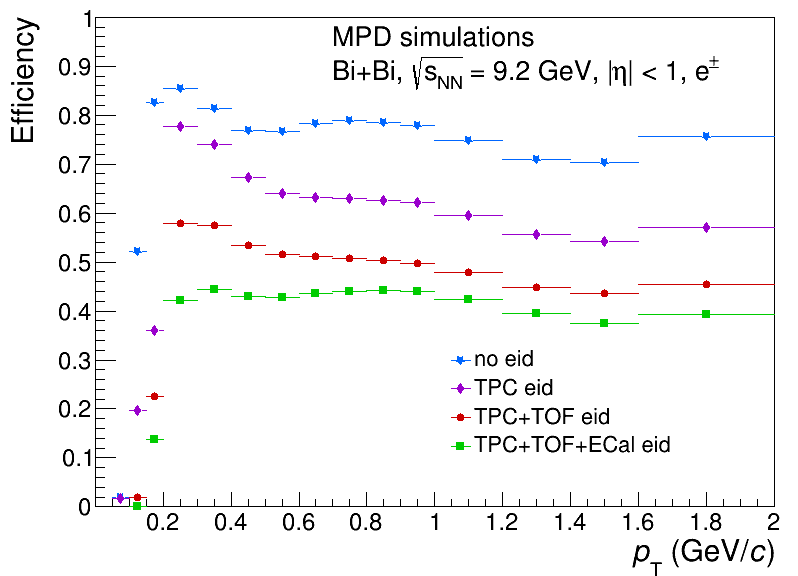}
\caption{Single-electron detection efficiency as a function of transverse momentum following a sequence of 1D cuts: fully reconstructed tracks without eid (blue), eid cut on TPC only (violet), on TPC+TOF (red), and on TPC+TOF+ECal (green).}
\label{fig_eff}
\end{figure*}

Figure~\ref{fig_eff} shows the single-electron detection efficiency as a function of transverse momentum obtained using sequential one-dimensional (1D) selection criteria applied separately to the TPC $dE/dx$, TOF $\beta$ and ECal $E/p$ observables.
The black lines in the left panel of figure~\ref{fig_pid2}  indicate the selection band of electrons in the TPC. In the TOF and ECal, electrons are selected by the requirements 
$|n_{\sigma}^{\beta}|< 2$ and $|n_{\rm \sigma}^{E/p}| < 3$, respectively. These cuts were chosen to maintain the electron-sample purity close to unity.  

\begin{figure*}
\centering
\includegraphics[scale=0.25]{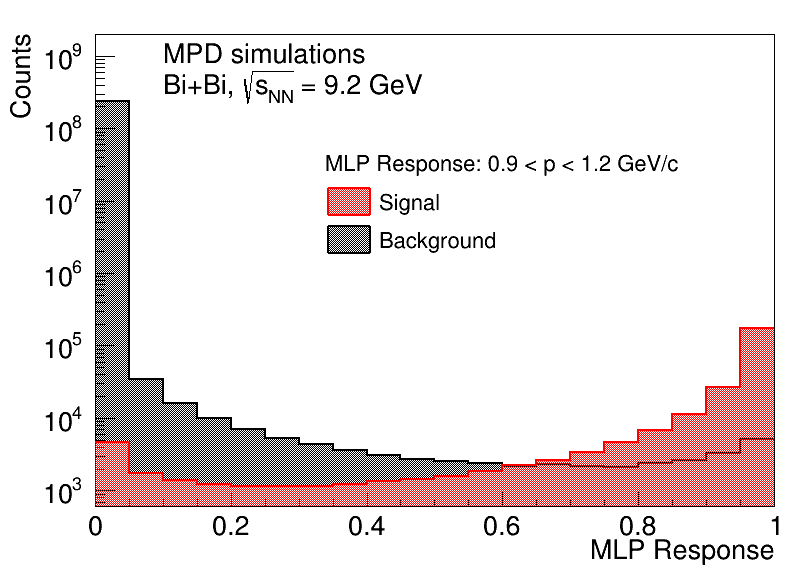}
\includegraphics[scale=0.24]{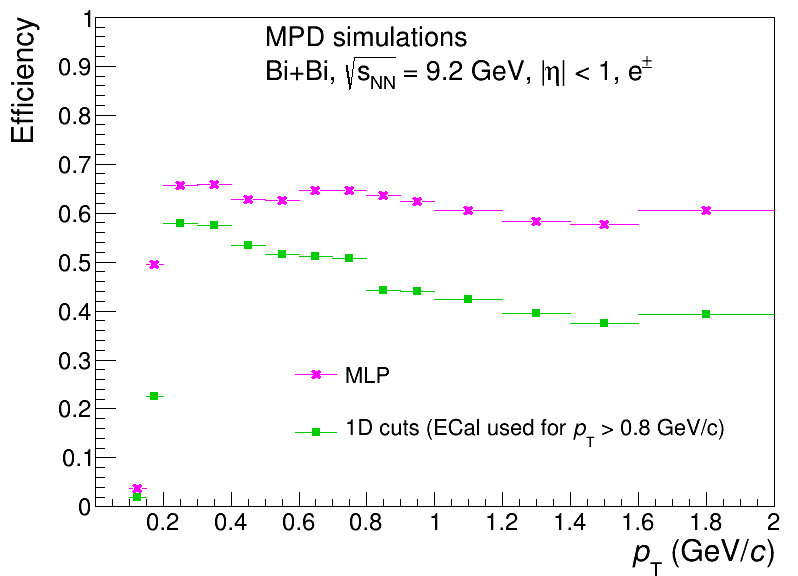}
\caption{Left panel: Response of the MLP classifier for signal ($e^{\pm}$) and background (hadrons) in the momentum interval, 0.9 $<$ $p$ $<$ 1.2 GeV/c. Right panel: Comparison of the single-electron detection efficiency based on the MLP approach (magenta)  and on the 1D cuts (green). The ECal information is used for $p_{\rm T} > 0.8$~GeV/$c$ in the cut-based method.}
\label{fig_ml}
\end{figure*}

Fully reconstructed tracks of charged particles, before applying any electron identification (``no eid''), are obtained with an efficiency of about 80\% over a wide transverse momentum range. The sequential application of 1D eid selections results in a gradual and significant reduction of the electron efficiency, reaching values of 40--45\% after applying the TPC, TOF and ECal cuts. 
The efficiency remains relatively flat in the range $0.3 < p_T < 1.0$~GeV/c and slightly decreases at higher momenta, where the separation power of the individual detectors becomes more limited. 
The gradual efficiency reduction 
reflects the stringent requirement to reject hadrons and achieve a high-purity electron sample (see subsection~\ref{sec_eid-perfC}) and illustrates the trade-off between efficiency and background rejection.

\subsection{Machine-learning-based identification}

Machine-learning (ML) techniques have recently been used for particle identification tasks 
~\cite{Papoyan:2025kdk,Graczykowski:2022zae,Karwowska:2024xqy}.  
These techniques can aid in finding an optimized selection,
thereby improving eid efficiency, without compromising electron purity. For the dielectron analysis presented below, we employ the MLP-based electron-identification method developed in ref.~\cite{Rode:2026mmj}. Its principal performance features relevant to the present study are summarized here for completeness and compared with the sequential 1D cuts.
 
We use the Toolkit for Multivariate Data Analysis (TMVA) package~\cite{TMVA,Brun:1997pa,Therhaag:2010zz,TMVA:2007ngy} from the CERN ROOT framework. We tested a neural network classifier, the Multi-Layer Perceptron (MLP),  and a tree-based machine learning algorithm, the Boosted Decision Tree (BDT).
The  MLP provided a slightly 
better performance and is therefore used throughout this work.  
 
The MLP is used here as a binary classifier to separate electrons from hadrons. Two statistically independent samples are used for training and testing, so that tracks from a given event appear in only one sample.  The training sample is further split into two sub-samples for training and validation in order to monitor the learning process and prevent overtraining. The model performance is evaluated using the independent test sample.
The machine-learning-based particle identification approach used in this analysis is described in detail in ref.~\cite{Rode:2026mmj}. 

Fully reconstructed tracks of charged particles satisfying the quality and matching requirements listed in 
table~\ref{tabI} are given as input to the model for training.  The electrons in this track sample (consisting of primary electrons and conversion electrons produced close to the primary vertex that are not rejected by the DCA cut) are classified as signal whereas all other tracks are treated as background. The electron-to-hadron ratio is kept equal to that in the UrQMD-generated events. 
A total of 11 input variables are used in the training: the reconstructed total momentum ($p$), pseudorapidity ($\eta$), azimuthal angle ($\phi$), number of hits in the TPC, truncated-mean specific energy loss in the TPC ($dE/dx$), velocity derived from the TOF measurement ($\beta_{\rm TOF}$), ($E/p$) ratio, velocity derived from the ECal timing measurement ($\beta_{\rm ECal}$), $\chi^{2}$ to the vertex ($\chi^{2}_{vertex}$),  longitudinal and transverse components of the distance of closest approach (DCA) to the primary vertex.

It was observed that training the model over the full momentum range leads to a degradation of the separation performance at high momentum. This effect is mitigated by dividing the dataset into momentum intervals
and training the model independently in each interval, resulting in a more uniform performance as a function of momentum. The left panel of figure~\ref{fig_ml} shows an example of signal and background separation for the momentum interval 0.9 $<$ $p$ $<$ 1.2 GeV/c. The threshold was chosen to match the purity of the complete 1D cuts in the independent validation sample.

The right panel of figure~\ref{fig_ml} shows the electron detection efficiency as a function of transverse momentum for the MLP-based identification in
comparison with the 1D cut-based approach. In the latter, the ECal $E/p$ cut is used only for tracks with p$_T >$ 0.8 GeV/c, since for lower p$_T$ values the hadron rejection is adequate with the TPC and TOF cuts only, as discussed previously (see figure~\ref{fig_pid2} middle panel). 
The MLP approach outperforms the cut-based selection over the entire momentum range. It provides a significant improvement in single-electron detection efficiency from $\sim40\%$ to $\sim60\%$ at high momentum, resulting in a gain of over a factor of 2 in dielectron efficiency (assuming the pair efficiency is given by the square of the single-electron efficiency).

\subsection{Electron sample purity}
\label{sec_eid-perfC}
\begin{figure*}
\centering
\includegraphics[scale=0.3]{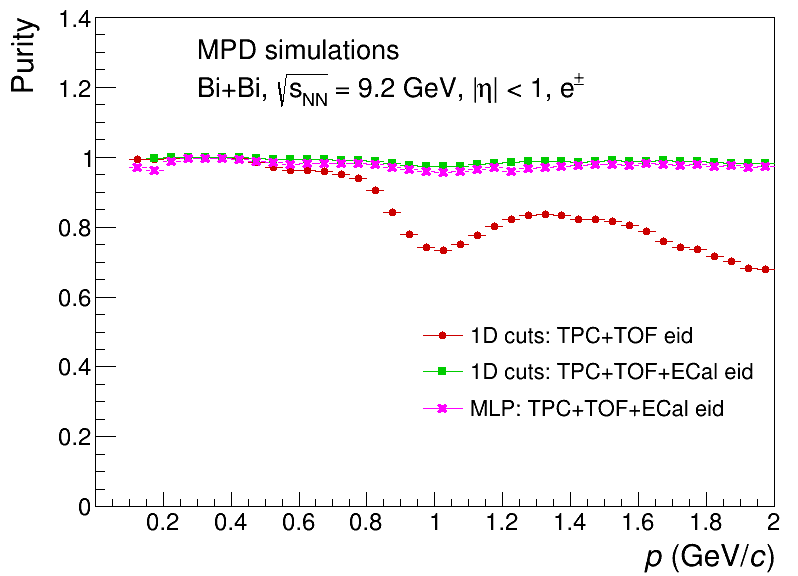}
\caption{Electron sample purity as a function of track  momentum for different identification strategies: TPC+TOF (red), TPC+TOF+ECal (green), and
MLP-based identification using all detectors (magenta).} 
\label{fig_purity}
\end{figure*}

The purity of the selected electron sample is shown in
figure~\ref{fig_purity} as a function of momentum for different identification strategies. The purity values are evaluated directly from the simulated sample without applying the inverse branching-ratio enhancement factors. The inclusive reconstructed electron sample is dominated by photon conversions and $\pi^{0}$ Dalitz decays, whose yields were not artificially enhanced; the enhanced rare decay channels therefore have a negligible effect on the total electron multiplicity and the  electron-sample purity.

The 1D cut-based approach using TPC and TOF information alone results in a significant contamination from hadrons, particularly at intermediate and high momentum, where the separation power of these detectors decreases. The inclusion of the ECal information leads to a substantial improvement in purity, yielding values close to unity over the entire momentum range. This reflects the strong discriminating power of the ECal $E/p$ observable for separating electromagnetic showers from hadronic interactions, particularly at intermediate and high momentum.

The MLP-based identification achieves a purity comparable to that of the full cut-based approach including the ECal, while maintaining a
significantly higher efficiency, as shown in the previous subsection. 
This demonstrates that the multivariate approach is able to optimally combine the detector responses and to achieve a superior balance between efficiency and purity compared to the 1D cut-based method. These improvements in eid performance are expected
to significantly enhance the sensitivity of dielectron measurements at NICA energies.  
Despite this improvement, dielectron measurements remain dominated by combinatorial background, as discussed in the following section.

\section{Challenges in dielectron  measurements}
\label{sec_challenges}
The measurement of dielectrons in heavy-ion collisions is particularly challenging due to the presence of a large combinatorial background.
Since the origin of the electrons is not known on a track-by-track basis, all reconstructed electron candidates are combined into pairs with all positron candidates from the same event. The resulting unlike-sign $e^{+}e^{-}$ invariant mass spectrum contains both the signal of correlated pairs and a large number of uncorrelated pairs. 

The physical signal ($S$) is obtained statistically by subtracting the combinatorial background ($B$) from the unlike-sign ($U$) spectrum:
\begin{equation}
    S(m) = U(m) - B(m)  
\end{equation}
$B$ is usually estimated using either mixed-event techniques or the like-sign pair combinations. In the present analysis, the combinatorial background is estimated using the geometric mean of the like-sign pair yields:
\begin{equation}
    B(m) = 2R(m)\sqrt{L_{++}(m)L_{--}(m)}
\end{equation}

where $R(m)$ accounts for possible differences in the unlike-sign and like-sign pair acceptance. Usually, $R(m)$ is determined experimentally using the mixed event technique. In the present simulation, we evaluated $R(m)$ directly using generator-level information (by comparing the truth combinatorial background in the unlike-sign spectrum to the geometric mean of the like-sign spectrum) and verified that $R(m) \simeq 1$ within the statistical uncertainties for all masses, for the detector configuration used in the present study. Its stability under the successive background-rejection steps is discussed in section~\ref{results}. In a real-data analysis, $R(m)$ would instead be determined from mixed events to account for charge-dependent acceptance due to local detector inefficiencies, inactive detector regions and other detector effects.

The challenge of the present dielectron analysis is to reduce the combinatorial background while preserving the signal as much as possible. The performance is quantified by two figures of merit, the signal-to-background ratio $S/B$ and the signal significance $S/\sigma_{S}$. 
 
The dominant sources of combinatorial background are incompletely reconstructed photon conversions and Dalitz decays of light neutral mesons, primarily $\pi^{0} \rightarrow \gamma e^{+}e^{-}$. 
These sources produce genuine correlated pairs. When only one of the two legs is fully reconstructed,  it forms random combinations with other uncorrelated electron tracks, thereby creating a large combinatorial background in the low-mass region. In addition, any residual hadron contamination further increases the background level.

\begin{figure*}
\centering
\includegraphics[scale=0.3]{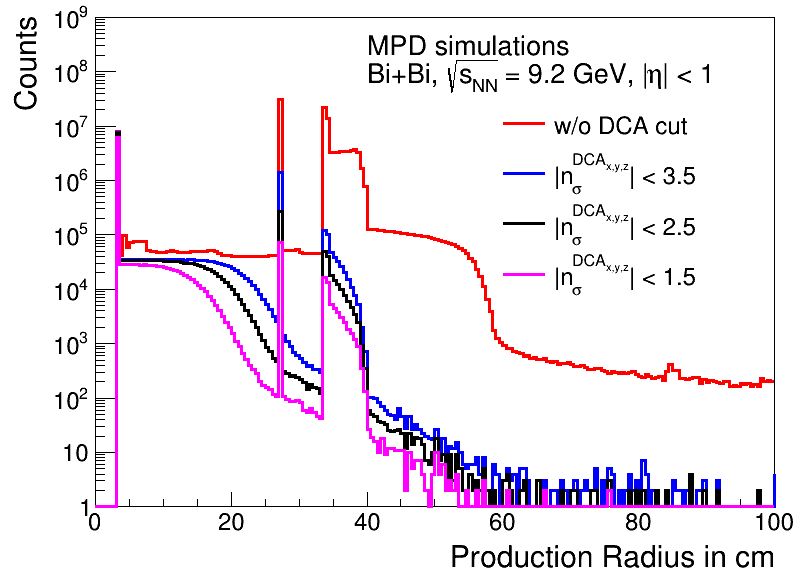}
\caption{Production radius distribution of electrons from photon conversions without and with various  DCA cuts.}
\label{fig_conv_rejection}
\end{figure*}

Photon conversions originate from secondary vertices and can therefore be suppressed by applying a cut on the DCA of the reconstructed tracks to the primary vertex as illustrated in figure~\ref{fig_conv_rejection}. The figure shows the production radius of electrons from photon conversions without and with various DCA cuts. 
The prominent yields of conversions occurring in the two inner vessels of the TPC (at radii of 27 and 34 cm) are clearly visible.  A DCA cut of 2.5$\sigma$ suppresses these conversions by more than two orders of magnitude, demonstrating the strong rejection power of this cut. However, conversions occurring close to the primary vertex, like those occurring in the beam pipe, at a production radius of $\approx$ 4 cm, remain largely unaffected by the DCA cut, since their reconstructed tracks are still compatible with originating from the primary vertex within the detector resolution. These remaining conversions are difficult to reject based solely on topological criteria and constitute a residual background component. After the DCA cut, the contributions from photon conversions and 
$\pi^0$ Dalitz decays are of comparable magnitude. 

\begin{figure*}
\includegraphics[scale=0.33]{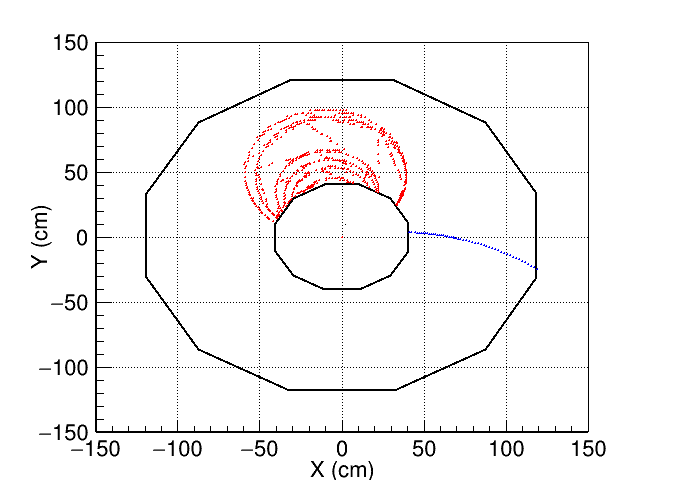}
\includegraphics[scale=0.33]{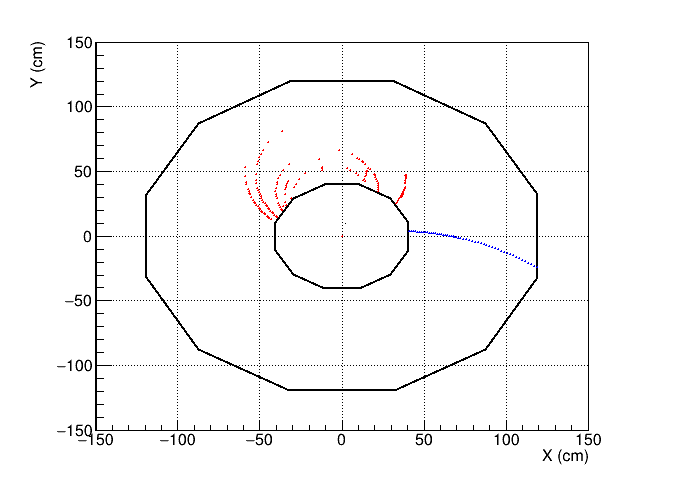}
\caption{Left: Hit distribution in the TPC pad plane of a high-$p_{\rm T}$ track (blue) and a low-$p_{\rm T}$ spiral track (red). Right: Reconstructed track segments obtained with the current tracking algorithm, demonstrating the incomplete reconstruction of the low-$p_{\rm T}$ spiral trajectory.}
\label{fig_spiraling_track}
\end{figure*}

A significant fraction of photon conversions and $\pi^{0}$ Dalitz decays have a high-$p_{\rm T}$ track and a low-$p_{\rm T}$ track.
The left panel of figure~\ref{fig_spiraling_track} illustrates the hit pattern in the TPC for such an event.
While the high-$p_{\rm T}$ track traverses the TPC in a relatively straight trajectory, the low-$p_{\rm T}$ electron follows a strongly curved trajectory and remains confined within the active volume of the detector. It undergoes multiple turns inside the TPC, producing a characteristic spiral pattern of hits, which complicates its reconstruction.

The right panel of figure~\ref{fig_spiraling_track} shows the corresponding tracks reconstructed by the current tracking algorithm. While the high-$p_{\rm T}$ track is correctly reconstructed, the low-$p_{\rm T}$ spiral track is only partially reconstructed. The current algorithm reconstructs tracks from inside to outside of the detector and thus the spiral track is reconstructed as
several short tracks or tracklets with alternating charge assignments.
In some cases, the track is not reconstructed at all, despite the presence of sufficient hits in the TPC.

This limitation leads to an incomplete reconstruction of $e^{+}e^{-}$ pairs from photon conversions and Dalitz decays. 
As a result, the fully reconstructed track remains in the electron sample and contributes to the combinatorial background, rather than being identified and removed at the pair level.

\section{Pair analysis strategy for background reduction}
\label{pair_analysis}

\subsection{Kinematic characteristics of background sources}

\begin{figure*}
\begin{center}
\includegraphics[scale=0.3]{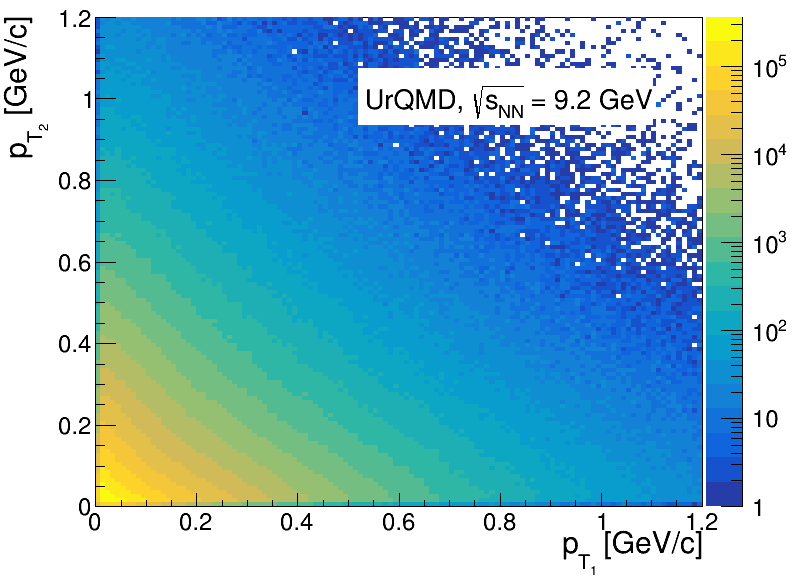}
\caption{Correlation between the transverse momenta of the two electrons from $\pi^{0}$ Dalitz decays using the UrQMD simulated Bi+Bi collisions at $\sqrt{s_{NN}} = 9.2$ GeV.} 
\label{pt_correl}
\end{center}
\end{figure*}

The electron pairs from the two main background sources, conversions and $\pi^0$ Dalitz decays, exhibit strong kinematic correlations (see figure~\ref{pt_correl}). 
A large fraction of pairs consists of one high-$p_{\rm T}$ electron, which is fully reconstructed,  and one low-$p_{\rm T}$ electron. The latter may not reach the outer TOF and ECal detectors, preventing its full reconstruction. 

For a fixed magnetic field, and within the central region used in this analysis, the minimum transverse momentum required for a charged particle to reach a given detector is practically pseudorapidity independent and is determined primarily by the detector radius.
At large pseudorapidity ($|\eta| \gtrsim 1.0$),  the transverse momentum threshold increases as particles with large longitudinal momentum leave the acceptance before reaching the outer detectors.
For the nominal magnetic field B = 0.5 T, 
particles with $p_{\rm T} \lesssim 32$~MeV/$c$ do not reach the TPC active volume and therefore leave no detectable signal. Particles with $32 \lesssim p_{\rm T} \lesssim 122 $ MeV/$c$ enter the TPC but do not reach the TOF detector. They produce hits only in the TPC, giving rise to the reconstruction of tracklets.   
Even if they do not reach the outer detectors and cannot be fully reconstructed, the partial information provided by these tracklets is useful and forms  the basis for the pair analysis method described below.

\begin{figure*}
\begin{center}
\includegraphics[scale=0.3]{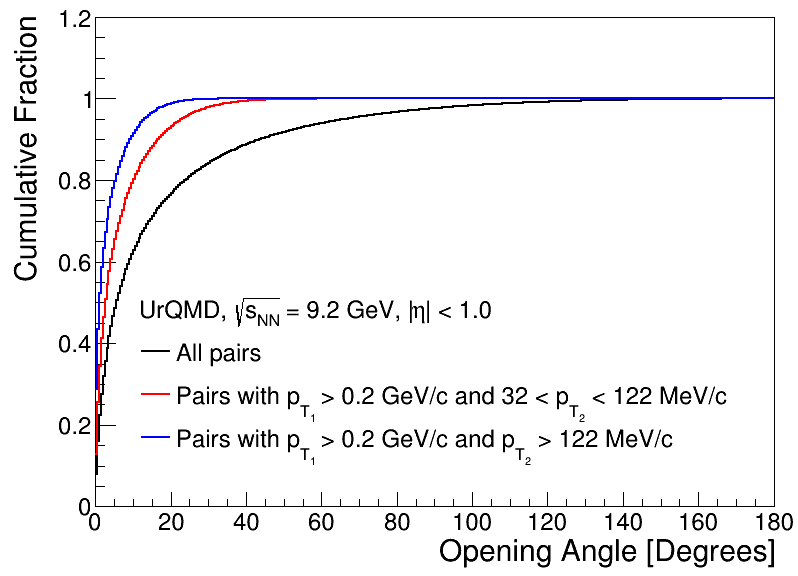}
\caption{Cumulative fraction of the 
opening angle distribution of dielectron pairs from $\pi^{0}$ Dalitz decays.  Black curve: all pairs; red curve: pairs with one track having ${p_{\rm T}}_1 > 200$~MeV/$c$ and the partner track with ${p_T}_{2}$ large enough to enter the TPC but not the TOF; blue curve: pairs with ${p_T}_1 > 200$~MeV/$c$ and 
${p_T}_{2}$ large enough to enter the TOF.}
\label{opening_angle} 
\end{center}
\end{figure*}

In addition, the background pairs are characterized by small opening angles. 
Figure~\ref{opening_angle} shows the cumulative distribution of the opening angle between the two electrons of $\pi^0$ Dalitz decays. The corresponding conversion-pair curves are very similar. It can be seen that about 95$\%$ of the pairs in which one of the electrons has $p_{\rm T} > $ 200 MeV/$c$ whereas the partner has 
$32 < p_{\rm T} < 122 $ MeV/$c$, have an opening angle $\theta < 20^{\degree}$.
For partners with $p_{\rm T} > 122 $ MeV/$c$, about 95\% of the pairs have an opening angle $\theta < 10^{\degree}$.

\subsection{Close TPC track cut}
The general concept of rejecting low-mass conversion and Dalitz pairs using fully and partially reconstructed electron candidates was introduced in the preliminary MPD study of ref.~\cite{MPD:2025jzd}. The present implementation develops this approach for strongly curved TPC tracklets, treating the partially reconstructed partner according to the detector information available for that track. 

The analysis strategy adopted in this work exploits the kinematic features discussed in the previous subsection to identify background pairs originating from conversions and $\pi^0$ Dalitz decays at the pair level. 
Explicitly, the analysis attempts to identify conversions and $\pi^0$ Dalitz pairs in which one electron is fully reconstructed whereas the second one is only partially reconstructed. A partially reconstructed track occurs mainly for low $p_{\rm T}$ particles ($32 \lesssim p_T \lesssim 122 $ MeV/$c$) producing tracklets as discussed above. It can also occur for high-$p_{\rm T}$ particles ($p_{\rm T} > 122$ MeV/$c$) due to inefficiencies of the matching to the outer detectors or inactive areas of these detectors.  In all cases, the partially reconstructed track provides both spatial information and a measurement of the specific energy loss ($dE/dx$), allowing partial particle identification. In addition, the fitted curvature provides a measure of the particle momentum. Due to the limited length of the tracklet for low $p_{\rm T}$ particles, the momentum resolution is reduced compared to fully reconstructed tracks. Nevertheless, the resulting invariant mass provides a useful discriminating variable for identifying pairs originating from photon conversions and $\pi^{0}$ Dalitz decays, through their characteristically low invariant mass.  
The method therefore combines complementary geometrical, particle-identification and kinematic information to identify conversion and Dalitz pairs.
 
Fully reconstructed electron tracks are paired with partially reconstructed ones and the resulting pairs are subject to the following selection, denoted close TPC track cut (CTC):
(i) small opening angle between the two trajectories,
(ii) compatibility of the $dE/dx$ signal of the partially reconstructed track with that of an electron, and
(iii) a small reconstructed invariant mass of the pair. The specific values of the angular and mass cuts are discussed in the next subsection.
Pairs satisfying these criteria are tagged as conversion or
$\pi^{0}$ Dalitz candidates and the tracks are removed from their pools of electron candidates.  
This approach allows the recognition of tracks originating from conversions and $\pi^0$ Dalitz even when only one of the tracks is fully reconstructed and enables the rejection of a significant fraction of tracks that would otherwise contribute to the combinatorial background.
In the following section, we quantify the performance of this method and evaluate the resulting improvement in the signal-to-background ratio.

\subsection{Pair analysis}
   
The reconstructed electron sample is divided into two pools. Pool 1 contains the fully reconstructed electron tracks, i.e. tracks satisfying the requirements listed in table~\ref{tabI} and the MLP eid selection.
Pool 2 contains partially reconstructed electron tracks, i.e. those that are reconstructed and identified in the TPC but either lack a valid match or failed identification in at least one of the two outer detectors, TOF or ECal.
These tracks are reconstructed
using slightly more relaxed requirements than those listed in table~\ref{tabI} for the fully reconstructed tracks. For example, the minimum number of TPC hits is reduced to 10, the acceptance is increased to $|\eta| \leq 1.2$ to allow finding the partner when the fully reconstructed track is close to the $|\eta| =1$ boundary. The track matching requirements to TOF or to  ECal, when applicable, are kept unchanged. One-dimensional eid cuts are used for the TPC $dE/dx$, the TOF $\beta$ and the ECal $E/p$.  

\begin{figure*}
\begin{center}
\includegraphics[scale=0.32]{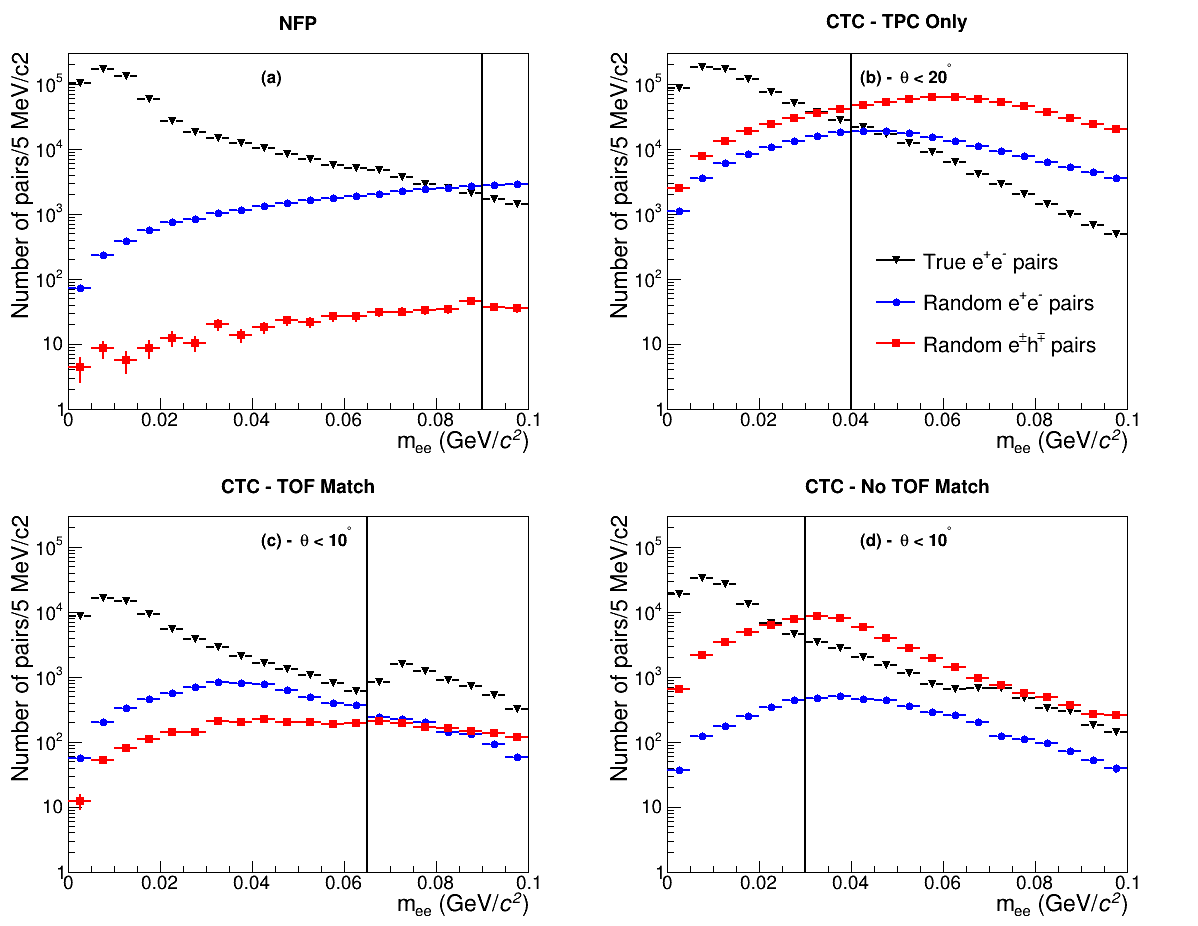}
\caption{Invariant-mass distributions of pairs removed at the different rejection stages. Black points show true $e^{+}e^{-}$ pairs from photon conversions and Dalitz decays, blue points show random $e^{+}e^{-}$ combinations, and red points show random electron-hadron combinations. The vertical lines indicate the applied invariant-mass cuts. The structure observed near 70 MeV/c$^2$ is associated in the simulation with residual conversions in the TPC entrance vessels.}
\label{fig_SB_CTC1}
\end{center}
\end{figure*}

The analysis is performed in three steps. In the first step, tracks from Pool 1 are paired with oppositely charged tracks from the same pool within the same event. Pairs with an invariant mass $m_{ee} < $ 90 MeV/$c^{2}$ are tagged as likely originating from $\pi^{0}$ Dalitz decays or photon conversions and the two tracks are excluded from any further pairing. This is denoted as the No Further Pairing (NFP) step. It was already used in the preliminary report in ref. \cite{MPD:2025jzd} and is retained here. 
To verify the validity of the assumptions behind this step, the information from the UrQMD generator is used. 
The top left panel of figure~\ref{fig_SB_CTC1} shows the invariant mass distribution of the pairs selected in the NFP step. 
The pairs are classified according to their origin: true $e^+e^-$ pairs, random $e^+e^-$, and random  $eh$ pairs. (The number of random $h^+h^-$ pairs is negligible). As expected, the removed pairs ($m_{ee} < $ 90 MeV/$c^{2}$)  are concentrated at low invariant masses and are dominated by true $e^+e^-$ pairs (96\% are true pairs) originating from photon conversions and $\pi^0$ Dalitz decays. The contribution from random combinations is comparatively small.

In the second step, the CTC is applied  to pairs formed by pairing the remaining tracks from Pool 1 with oppositely charged tracks from Pool 2. 
Three cases are distinguished according to the detector information available for the partially reconstructed partner:

a) partner tracks detected only in the TPC.
These are mostly partner tracks with ${p_T}_2 < 122 $ MeV/$c$. 
In this case, pairs with opening angle $\theta < 20^{\degree}$ and invariant mass  $m_{ee} < 40 $ MeV/$c^{2}$ are tagged as potential $\pi^{0}$ Dalitz or conversion pairs and the two tracks are removed from their pools. The angular cut is motivated by the cumulative distributions shown in figure~\ref{opening_angle} whereas the mass cut is derived from figure~\ref{fig_SB_CTC1}b
that shows the invariant-mass distribution and the origin of  the pairs selected in this case
after applying an angular cut of $\theta < 20^{\degree}$. Although the momentum of the partner track is reconstructed only from a tracklet, 
the selected sample with a mass cut of $m_{ee} < 40 $ MeV/$c^{2}$ is still dominated by genuine conversion and $\pi^0$ Dalitz pairs. Approximately 75\% of the selected pairs correspond to true $e^+e^-$ pairs, while the remaining contribution originates predominantly from random electron-hadron combinations.

b) partner tracks detected in the TPC and the TOF. These are mostly tracks with ${p_T}_2 > $122 MeV/$c$. 
Following figure~\ref{opening_angle}, the angular cut is tightened to  $\theta < 10^{\degree}$ whereas the mass cut is derived from figure~\ref{fig_SB_CTC1}c 
that shows the mass distribution and the origin of the pairs in this case after applying the angular cut.
Here, the additional particle identification provided by the TOF allows the mass cut to be set at a higher value. 
A cut of $m_{ee} < $ 65 MeV/$c^{2}$ ensures that the fraction of true pairs is about 89\%. 
Pairs satisfying these criteria ($\theta < 10^{\degree}$ and $m_{ee} < 65 $ MeV/$c^{2}$) are tagged as potential $\pi^{0}$ Dalitz or conversion pairs and the two tracks are removed from their pools.

c) partner tracks detected in the TPC and ECal but not in the TOF. 
These are mainly tracks with ${p_T}_2 > 150 $ MeV/$c$ that lack a matched TOF hit, predominantly because of inactive areas or matching inefficiencies.
In this case, the angular cut is $\theta < 10^{\degree}$, the same as in case b) and the mass cut is set at $m_{ee} < 30 $ MeV/$c^{2}$, following figure~\ref{fig_SB_CTC1}d. Information from the UrQMD generator reveals that with these cuts the fraction of true pairs is about 80\%. As in the previous two cases, pairs satisfying these criteria are tagged as potential $\pi^{0}$ Dalitz or conversion pairs and the two tracks are removed from their pools. 
 
One should note that in the very rare cases where more than one track satisfies the above pairing selections, the result does not depend on the order in which pairs are processed. Furthermore, the fraction of signal tracks accidentally removed in any of the above cases is relatively low compared to the benefit of background reduction as will be quantified in the next section.

The combination of the NFP and the three CTC categories yields a total sample of tagged pairs, 83\% of which are true pairs.
Overall, figure~\ref{fig_SB_CTC1}
demonstrates that the CTC predominantly selects genuine conversion and Dalitz pairs, allowing the recognition of these pairs even when only one track is fully reconstructed. It validates both the physical assumptions underlying the CTC and the invariant-mass cuts used in the analysis.

In the third step, the remaining Pool 1 tracks  with transverse momentum $p_{\rm T} > 200$ MeV/$c$, within the same event, are paired to construct the unlike-sign ($U$) and like-sign ($L_{++}$ and $L_{--}$) invariant mass spectra. The reconstructed signal is determined as described in section~\ref{sec_challenges}. 

We note that the artificial enhancement of rare dielectron sources could, in principle, also modify the probability of accidental NFP and CTC associations. This effect is not exactly removed by the inverse branching-ratio factors but is expected to be small in the present analysis because the electron multiplicity is dominated by photon conversions and $\pi^{0}$ Dalitz decays, whose rates were not enhanced. The enhanced sources contribute only a few percent to the total electron multiplicity. Their contribution to the probability of accidental associations is therefore correspondingly small.
 
\begin{figure*}
\begin{center}
\includegraphics[scale=0.25]{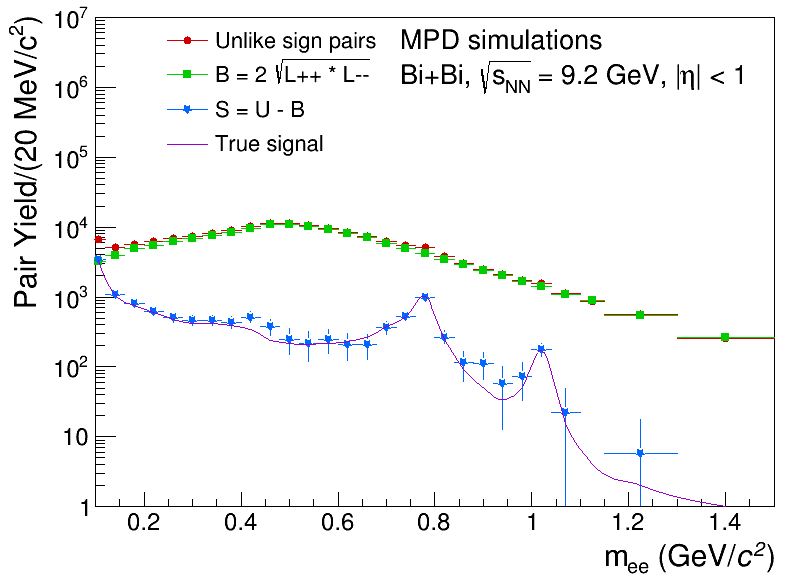}
\includegraphics[scale=0.25]{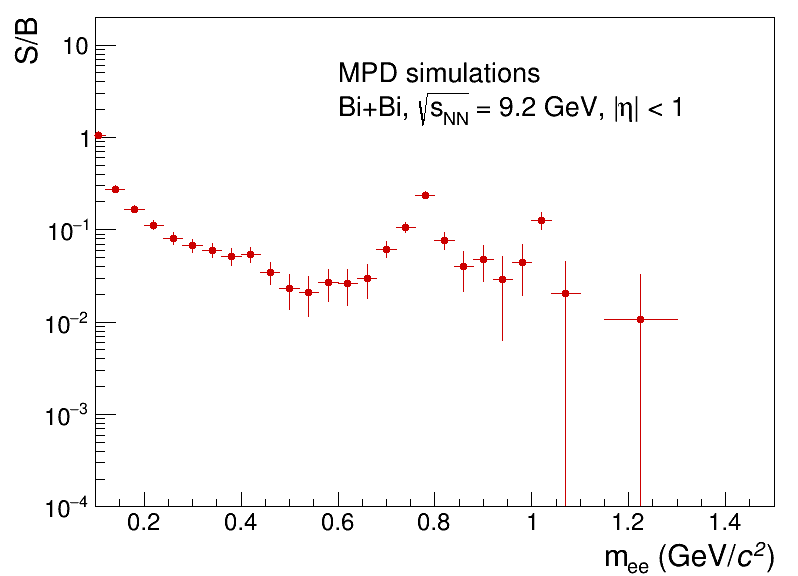}
\caption{ Left panel: invariant mass distributions of the unlike-sign pairs U (dots), combinatorial background B estimated from the geometric mean of the like-sign pairs (squares), reconstructed signal S = U-B (stars) and true reconstructed signal obtained from generator information (line). Right panel: differential signal-to-background ratio after applying the full pair-analysis strategy.}
\label{fig_SB_CTC}
\end{center}
\end{figure*}

\section{Results and discussion}
\label{results}
In this section, we quantify the performance of the pair-analysis strategy described in the previous section.
The central values are presented after applying downscaling factors to each pair to restore the physical branching ratios of the dilepton sources (see section~\ref{sec_simul}). The quoted uncertainties, $\sigma_S$, are the statistical uncertainties of the signal that would be obtained in a projected measurement with the same $U$ and $B$ yields, namely $\sigma_S = \sqrt{ U+B}$.
Figure~\ref{fig_SB_CTC} shows the invariant-mass distributions after applying the full
analysis strategy. The left panel presents the unlike-sign spectrum ($U$), the
like-sign estimate of the combinatorial background ($B$), the reconstructed signal ($S=U-B$), and the true reconstructed signal obtained from the generator information. The reconstructed signal is consistent with the true signal within the available statistical precision, indicating  that the like-sign background estimate and the rejection procedure do not introduce a significant distortion of the signal shape. The right panel shows the corresponding differential signal-to-background ratio as a function of invariant mass.

The effect of the different rejection stages is summarized in table~\ref{tab2}. The quantities reported in the table are evaluated in the mass interval $0.2 < m_{ee} < 0.7$~GeV/$c^{2}$.
Before applying any pair rejection, the starting value of the signal-to-background ratio is about $1.3\%$. The NFP step reduces the combinatorial background by about 36\% while retaining the signal within the statistical uncertainties. The subsequent application of the CTC step further reduces the background, reaching a total reduction by a factor of about 4.0 with respect to the starting value.  
At the same time, most of the reconstructed signal is preserved, resulting in an increase of the signal-to-background ratio to
approximately $4.5\%$, corresponding to an improvement by a factor of about 3.5 compared to the baseline. The statistical significance
also improves, from 8.2 to 14.4. These results demonstrate that the
CTC method removes the combinatorial background more efficiently than the signal and therefore improves the statistical reach of the measurement.

\begin{table*}[ht]\centering
\vglue4mm
\setlength{\tabcolsep}{10pt} 
\renewcommand{\arraystretch}{1.2} 
\begin{tabular}{cccc} \hline\hline
\textbf{Quantity} & \textbf{Starting values}&\textbf{NFP}&\textbf{NFP+CTC}\\
\hline
$U$    &   839191   &543127  &    217403 \\
$B$    &   828564   &532072   &   208009\\
$S = U-B$ & 10627$\pm$1291  &11055$\pm$1037   &    9394$\pm$652 \\
$S/B$ ($\%$)&   1.3 & 2.1 &4.5 \\
Significance& 8.2    & 10.7 & 14.4 \\

\hline
\hline
\end{tabular}
\caption{Results evaluated in the invariant-mass range $0.2 < m_{ee} < 0.7$~GeV/$c^2$ at the different stages of the analysis
after applying downscaling factors to each pair as discussed in section~\ref{sec_simul}. $U$ is the unlike-sign yield, $B$ is the combinatorial background approximated by the geometric mean of the like-sign yield, and $S$ is the reconstructed signal.
The significance is evaluated as 
$S/\sigma_S$, where $\sigma_S$ is the statistical uncertainty in $S$ that would be obtained in a real measurement of the same $U$ and $B$ yields, namely $\sigma_S = \sqrt{\rm U+\rm B} $.}
\label{tab2}
\end{table*}

\begin{figure*}
\begin{center}
\includegraphics[scale=0.32]{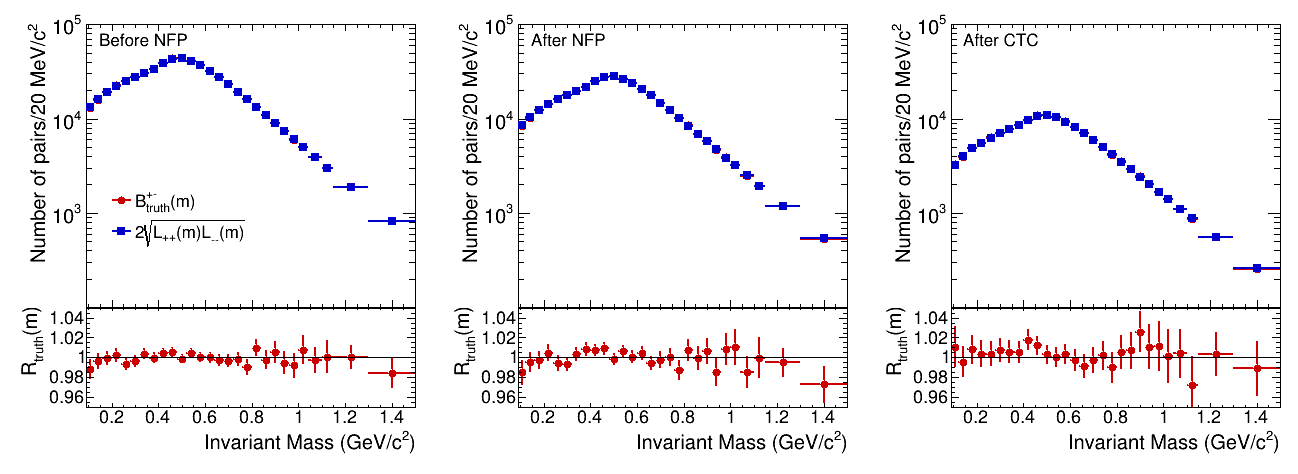}
\caption{Comparison of the true combinatorial background ($B_{\rm truth}^{+-}(m)$) and the geometric mean of the like-sign pairs ($2\sqrt{L_{++}(m)L_{--}(m)}$) for the three analysis steps, namely, before NFP, after NFP and after NFP+CTC. Their ratio, $R_{\rm truth}(m)$, is shown as a function of invariant mass in the bottom panels.}
\label{fig_R}
\end{center}
\end{figure*}

As an additional check of the like-sign background estimate, the factor $R_{\rm truth}(m)$ is evaluated directly from generator truth as: 
\begin{equation}
    R_{\rm truth}(m)=\frac{B_{\rm truth}^{+-}(m)}
{2\sqrt{L_{++}(m)L_{--}(m)}}
\end{equation}
Figure~\ref{fig_R} shows this ratio for the initial sample before NFP, after NFP, and after the full NFP+CTC rejection procedure. It remains consistent with unity within the statistical uncertainties at all three stages of the analysis. This demonstrates that the successive pair selections do not introduce a significant relative acceptance difference between unlike-sign and like-sign combinations in the present simulation. As mentioned earlier, for real data, $R_{\rm truth}(m)$ will be determined using mixed events to account for any charge-dependent acceptance effects.

In order to assess the robustness of the quoted performance, the principal selection thresholds were varied around their nominal values. The invariant-mass thresholds used in the NFP and CTC selections were varied by $\pm10$ MeV/c$^2$, while the CTC opening-angle requirement was varied by $\pm 5^{\degree}$ for TPC-only partner tracks and by $\pm 2^{\degree}$ for the TPC+TOF and TPC+ECal categories. For each variation, the relative changes in the reconstructed signal S, the signal-to-background ratio S/B, and the statistical significance were evaluated with respect to the nominal result. The deviations obtained from the individual cut variations were then added in quadrature. The combined variations were found to be 8.6\%, 11.1\%, and 3.3\% for S, S/B, and the statistical significance, respectively. The relatively small combined variation in significance indicates that the improvement in statistical sensitivity is robust against reasonable changes of the selection thresholds. This study is intended as a test of the stability of the analysis procedure, not as a complete evaluation of systematic experimental uncertainties.

The performance values quoted here are averages over the minimum-bias event sample. Since the combinatorial background, tracking reconstruction, and eid performance depend on event multiplicity, the rejection efficiency and the resulting S/B improvement may vary with collision centrality. A reliable centrality-resolved performance evaluation requires a dedicated analysis with sufficient statistics in each centrality class and is beyond the scope of the present work.

The present implementation is limited by the efficiency and quality of the reconstruction of low-$p_{\rm T}$ tracklets. 
The analysis of the final tracks remaining in Pool 1, using generator-level information, reveals that about 55\% of them originate from 
$\pi^0$ Dalitz decays or photon conversions 
and have a partner with 
$p_{\rm T} > 32 $ MeV/c, i.e. a partner that reaches the TPC active volume but is not reconstructed by the current Kalman-filter tracking algorithm.
Further improvements in the reconstruction of strongly curved low-momentum tracks should therefore enhance the recognition and rejection of incompletely reconstructed conversion and Dalitz pairs and bring the performance closer to the full potential of the proposed strategy.

\section{Summary}
\label{summ}

We have studied the performance and capabilities of the MPD detector for dielectron measurements at NICA energies using simulated $^{209}$Bi+$^{209}$Bi collisions at $\sqrt{s_{NN}}=9.2$~GeV. Electron identification is achieved using the combined information from the TPC, TOF and ECal detectors. The MLP-based electron-identification method of ref.~\cite{Rode:2026mmj} is used as part of the present dielectron analysis and provides a higher electron efficiency than the sequential 1D reference selection at comparable purity.

A pair-analysis strategy was developed to reduce the
combinatorial background arising from incompletely reconstructed photon conversions and $\pi^{0}$ Dalitz
decays. The method exploits partially reconstructed tracks and combines the pair opening-angle, the TPC $dE/dx$ signal and an approximate reconstruction of the pair invariant mass to identify and remove background tracks at the pair level.

With the current reconstruction algorithm, the outlined strategy reduces the  background by a factor of 4.0, while retaining approximately 88\% of the signal, resulting in an improvement of the signal-to-background ratio by a factor of about 3.5 
in the mass range $0.2 < m_{ee} < 0.7$~GeV/$c^2$. The quoted performance corresponds to minimum-bias collisions; its centrality dependence remains to be studied separately.

The MLP-based electron identification and the CTC approach provide complementary benefits. The MLP selection increases the signal efficiency while preserving high electron purity whereas the CTC approach improves the signal-to-background ratio by suppressing tracks from background sources before the final pairing. Together, these two elements significantly improve the projected sensitivity of the MPD detector for dielectron measurements at NICA energies. Further improvements in low-$p_{\rm T}$ track reconstruction in the TPC can provide additional background rejection.

\acknowledgments
We thank the MPD Collaboration for the encouragement and support. We also thank the staff of the computing center of 
the Meshcheryakov Laboratory of Information Technologies at the Joint Institute for Nuclear Research for providing the computing resources and technical support for this research. This work was funded by the Ministry of Science and
Higher Education of the Russian Federation, Project “Fundamental
and applied research at the NICA (JINR) megascience experimental complex” FSWU-2025-0014 and by the Shandong Provincial Natural Science Foundation (Project No. ZR2024QA192).

\bibliographystyle{JHEP}

\bibliography{jinst}

\end{document}